# Do Bi-Stable Steric Poisson-Nernst-Planck Models Describe Single Channel Gating?

Nir Gavish, Chun Liu, and Bob Eisenberg

Department of Mathematics, Technion – Israel Institute of Technology, Haifa, Israel
Department of Applied Mathematics, Illinois Institute of Technology, Chicago, USA
Department of Physiology and Biophysics, Rush University, Chicago and Department of Applied Mathematics, Illinois Institute of Technology, Chicago, USA

E-mail:
ngavish@technion.ac.il
bob.eisenberg@gmail.com
cxl41@psu.edu

Abstract

Experiments measuring currents through single protein channels show unstable currents, a phenomena called the gating of a single channel. Channels switch between an 'open' state with a well defined single amplitude of current and 'closed' states with nearly zero current. The existing mean-field theory of ion channels focuses almost solely on the open state. The physical modeling of the dynamical features of ion channels is still in its infancy, and does not describe the transitions between open and closed states, nor the distribution of the duration times of open states. One hypothesis is that gating corresponds to noise-induced fast transitions between multiple steady (equilibrium) states of the underlying system. In this work, we aim to test this hypothesis. Particularly, our study focuses on the (high order) steric Poisson-Nernst-Planck-Cahn-Hilliard model since it has been successful in predicting permeability and selectivity of ionic channels in their open state, and since it gives rise to multiple steady states. We show that this system gives rise to a gating-like behavior, but that important features of this switching behavior are different from the defining features of gating in biological systems. Furthermore, we show that noise prohibits switching in the system of study. The above phenomena are expected to occur in other PNP-type models, strongly suggesting that one has to go beyond over-damped (gradient flow) Nernst-Planck type dynamics to explain the spontaneous gating of single channels.

Graphical Abstract

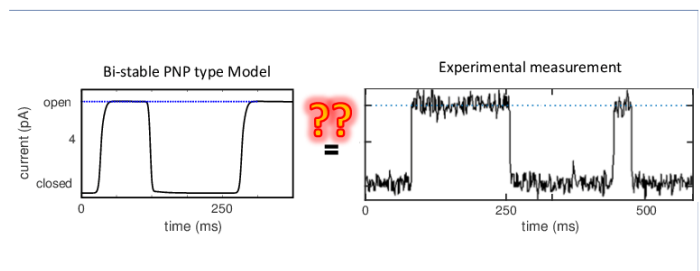

# 1   Introduction

Ion channels are protein molecules that conduct ions (such as Na$^+$, K$^+$, Ca$^{2+}$, and Cl$^-$ that might be named bioions because of their universal importance in biology) through a narrow pore of fixed charge formed by the amino acids of the channel protein. Membranes are otherwise quite impermeable to natural substances, so channels are gatekeepers for cells and act as natural nano-valves. Controlled ion permeation through ion channels is one of the most important living processes [21, 53], governing an enormous range of biological function in health and disease [2].

Ion channels have been studied one at a time for nearly forty years in a triumph of experimental science. For an overview of these efforts, see, e.g., the book of Nobel laureates Sakmann and Neher [41]. Measurements are now commonplace. They are made in thousands of laboratories every week for hundreds of types of channels.

But the commonplace has hidden the obvious. Single channels are unstable devices. They stochastically switch between two current levels, in a process called gating. One current level is the main conductance state, and the second current level is nearly zero, corresponding to a closed channel. Some sub-conductance states are seen as well. Gating in a wide range of ionic channels has specific well-known characteristics, observed and studied in thousands of papers [41].

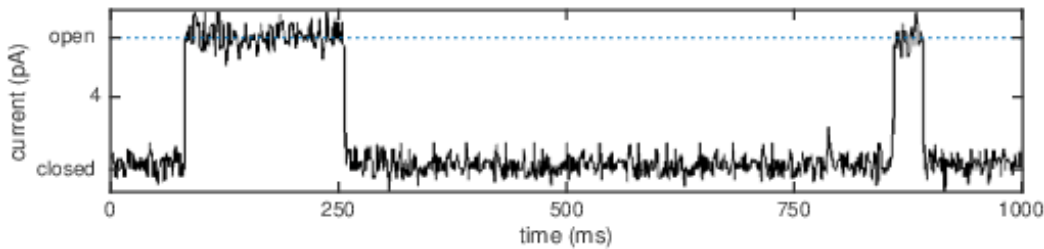

Figure 1: Measurement of current vs time of an isolated RyR channel [21, 41]. The current switches abruptly between a level close to zero to an open level (~7.3 pA). The current remains at an open level for a stochastic duration. Significantly, the level of current once the channel is open is independent of time, and no intermediate values of current are observed (but see remark regarding the epi-phenomena of 'subconductance states'). Image adapted from [21]

The current vs time switches between a level close to zero to an open level (very different for different types of channels, from 1 to 100 pA), see, e.g., Figure 1. Switching is abrupt, faster than one or two microseconds. While the channel is in the process of opening, bizarrely diverse behavior is observed, resembling the trajectories of Brownian motion [37, 48], but once the channel is open, the mean current is independent of time, and the open channel noise is well behaved [20, 43, 44]. Channel opening occurs at stochastic times: the duration of the channel opening is stochastic. Closed and open duration histograms are often nearly single exponential functions.

The universal phenomena of single channel gating, observed in so many types of channels with such different structures, can be described (in all likelihood) by a single model that depends on one main process that is common to all channels and does not depend on specific properties of each channel. Most theoretical studies of gating consider Markov Models or kinetic models, see, e.g., [41, 53, 47] and references within. While these models describe the dependence of single channel openings (duration, open probability, and occasionally latency) on time in *one set of conditions*, the models do not describe dependence on concentrations or electrical potentials, nor do they conserve current [8]. In what follows, we consider continuum models for ion channels derived by the energetic variational (EnVarA) approach that allows a self- consistent description of the system, while conforming with physical conservation laws, see for example[16, 23]. Self consistent means that all variables satisfy all differential equations and boundary conditions with one set of parameters under all experimental conditions. Non-transferrable models are examples of inconsistent descriptions that require different parameters under different conditions. Markov models of single channel gating (and much) else are often inconsistent because their parameters change as experimental conditions (voltages or ion concentrations) change. Poisson-Nernst-Planck models of semiconductors [42, 51], and Fermi-Poisson models of open calcium channels are examples of consistent models [30, 32] because they fit data under a range of conditions with one set of parameters. Continuum mean-field theories of electrolytes, which are generalizations of Poisson-Nernst-Planck (PNP) models, have been widely used in studies of ion channels during the last two decades, for reviews see [10, 18, 4] and references within. Correlations introduced by the finite-size of the ions play a crucial role in determining the permeability and selectivity of ion channels. Accordingly, PNP equations with steric effects, and in particular the PNP-steric model [23], have been successful in predicting permeability and selectivity of ionic channels in its open state. Real channels, however, are closed a substantial fraction of their time and their switching behavior is an important determinant of biological function. Nevertheless, the existing models focus almost solely on characterization of open channels [10, 18, 4]. Namely, they consider a conditioned experimental system in which the channel is open and conducting current. Thus, notwithstanding recent advances [34, 49], theoretical modeling of the dynamical features of ion channels is still in its infancy, and does not describe the transitions between open and closed states, nor the distribution of the duration times of open states.

The Poisson-Nernst-Planck (PNP) equation itself and a wide family of generalized PNP equations give rise to a unique steady-state [15], and therefore are unlikely to be able to describe gating without the addition of specific time dependent features, like a voltage sensor [24]. In contrast, PNP equations involving multiple regions of piecewise constant permanent charge were shown to give rise to multiple steady states [7, 33]. Some of these solutions were found to be stable, giving rise to a bi-stable model [5]. Similarly, the PNP-steric equation with spatially constant permanent charges (which successfully describes the permeation in an open current-conducting single channel [23]) was shown to give rise to multiple steady states [29]. These solutions, however, turned out to be unstable, and in fact reveal that the PNP-steric equation is ill-posed in the regime of high ionic concentrations where it gives rise to multiple steady-states [14]. This work [14] led to a PNP equation with high-order steric effects (also known as the steric PNP-Cahn-Hilliard or PNP-CH model) which is well-posed at high ionic concentrations and, furthermore, gives rise to bi-stable behaviour [14]. The fundamental question facing all these investigators was [9]: *Does gating correspond to noise-induced fast* transitions between multiple steady (equilibrium) states of the underlying system? In this study, we see if a class of models describes the gating phenomena seen in a vast number of single channel experiments. Particularly, we address the question whether switching behavior in PNP equations with high-order steric effects has the universal features of gating in biological ion channels.

The paper is organized as follows: We first present the Poisson-Nernst-Planck-Cahn-Hilliard (PNP-CH) model for ion channels, and focus on its steady states and their stability. In particular, as expected, we show that the equation describes a bi-stable model which reflects a competition between multiple cationic species inside channel. In the preceding section, we consider the dynamics of the bi-stable model. We show that the model can describe a switching behavior, albeit under rather specific conditions, and in the absence of noise. Unexpectedly, we find that the introduction of noise prohibits switching. A detailed comparison between the switching behavior of the model, and gating behavior in biological ion channels, shows that the switching behavior described by the model does not have the defining features of gating in biological systems. We are, therefore, drawn to conclude that the hypothesis that gating phenomena is described by noise-induced fast transitions between multiple steady (equilibrium) states of PNP-type equations is incorrect. Possible alternative paradigms for gating are presented in the Discussion section.

## 2 Mathematical Model

We follow the unified energetic variational framework for ionic solutions, formulated by Liu, more than anyone else, that treats ions as solid charged spheres of finite size [40, 39, 11, 25, 23, 12, 52]: The functional formThe functional form $\mathcal{A}$ reduces to the Helmholtz free energy when $\phi$ satisfies Poisson's equation (3). describing a system with $N$ species with concentrations $c_i$ and valences $z_i$, where $i = 1, \cdots N$, is given by

$$\mathcal{A}(c, \phi) = \int_\Omega k_B T \underbrace{\sum_{i=1}^N c_i \left( \ln \frac{c_i}{\bar{c}_i} - 1 \right)}_{entropy} + \underbrace{q \left( z^T c + \rho_0(x) \right) \phi - \frac{\varepsilon}{2} |\nabla \phi|^2}_{electrostatic} + \underbrace{\psi(c; a)}_{\text{Lennard–Jones}} \, dx \qquad (1)$$

where $\phi$ is the electric potential, $c = (c_1, \cdots, c_N)$, $z = (z_1, \cdots, z_N)$, $q$ is the unit of electrostatic charge, $k_B$ is Boltzmann's constant, $T$ is the temperature, $\varepsilon$ is the relative dielectric constant, assumed to be uniform, and $\rho_0(x)$ is permanent charge. Energy of the repulsion of the ions, treated as solid spheres of size $a = (a_1, \cdots, a_N)$, is included in $\mathcal{A}$ as Lennard-Jones forces. The Lennard-Jones potential term $\psi(c, a)$ is a convolution integral with a singular kernel that imposes analytical and numerical difficulties. These difficulties are particularly unfortunate because there is only historical, not physical justification for using the Lennard Jones formulation of interparticle interactions. The combining rules for particles of different diameter are particularly hard to justify. We note that recently Liu, Xie and Eisenberg [32] have used a Yukawa formulation that works around some of these problems. Approximation of the Lennard-Jones potential by a band-limited function gives rise to the local approximation of the form [23, 28, 29]

$$\psi(c, a) = \frac{1}{2} c^T G c + \frac{1}{2} (\nabla c)^T \Sigma \nabla c + \cdots, \qquad (2)$$

where $G$ and $\Sigma$ are $N \times N$ symmetric matrix with positive entries $g_{ij}$ and $\sigma_{ij}$, respectively. The papers [23, 28, 29] focused on the leading order approximation of the Lennard-Jones potential $\psi(c,a)$ (namely $\psi \approx \frac{1}{2} c^T G c$). The resulting equation, however, turned out to be ill-posed for highly concentrated electrolytes in relevant parameter regimesThese parameter regimes are associated with multiple steady-state solutions, and hence are the ones relevant to this study. [14]. Therefore, it is necessary to also account for high-order steric effects ($\Sigma \neq 0$) to remedy the ill posed nature of earlier analyses. In what follows, we consider the case $\Sigma = \sigma I$ which is sufficient to assure that the resulting equation is well-posed [14].

The electric field is required to be a critical point of $\mathcal{A}$, yielding Poisson's equation

$$\frac{\delta \mathcal{A}}{\delta \phi} = \varepsilon \Delta \phi + q \, (z^T c + \rho_0) = 0. \tag{3}$$

Taking the evolution of the ionic species results in the form of a Nernst-Planck type equation yields

$$\begin{aligned}\frac{dc_i}{dt} &= \frac{1}{k_B T} \nabla \cdot \left( D_i(x) c_i \nabla \frac{\delta \mathcal{A}}{\delta c_i} \right) \\ &= \nabla \cdot \left[ D_i(x) \left( \nabla c_i + \frac{c_i}{k_B T} \left( z_i q \nabla \phi + \sum_{j=1}^{N} g_{ij} \nabla c_j - \sigma \nabla \Delta c_i \right) \right) \right], \quad i = 1, \cdots, N, \end{aligned} \tag{4}$$

where $D_i(x)$ is the diffusion coefficient of the $i^{th}$ species. No-flux boundary conditions are implemented for both the ionic concentration and potential at the side walls (orthogonal to the direction of current flow). The arising *steric Poisson-Nernst-Planck-Cahn-Hilliard* (steric PNP-CH) system accounts for the various interactions between the components in a self-consistent way, while satisfying the second law of Thermodynamics, via Onsager's relation,

$$\frac{d\mathcal{A}}{dt} = -\frac{1}{k_B T} \int_\Omega \sum_{i=1}^{N} D_i(x) \left| \nabla \frac{\delta \mathcal{A}}{\delta c_i} \right|^2 dx \leq 0.$$

We consider the non-dimensional variables

$$\tilde{\phi} = \frac{q}{k_B T} \phi, \quad \tilde{c}_i = \frac{c_i}{\bar{c}_i}, \quad \tilde{x} = \frac{x}{\lambda}, \quad \tilde{t} = \frac{D_i(0)}{\lambda^2} t,$$

where $\bar{c}_i$ is the bath concentration of species $i$. For simplicity, we assume here that the left and right bath concentrations are equal, i.e., $\bar{c}_i = \bar{c}$ for $i = 1, 2, \cdots, n$. The corresponding non-dimensional parameters are

$$\tilde{g}_{ij} = \frac{\bar{c}_i}{k_B T} g_{ij}, \quad \tilde{\sigma} = \frac{\bar{c}_i}{\lambda^2 k_B T} \sigma, \quad \tilde{D}_i = \frac{D_i}{D_i(0)}, \quad \tilde{\rho}_0 = \frac{\rho_0}{\bar{c}}.$$

Note that this scaling is standard, see, e.g.[23], except that time is scaled by the characteristic diffusion coefficient of an ion *inside* the channel as $O(1)$, rather than by the bulk diffusion coefficient. We note that despite the use of the usual scaling for time (namely, the diffusion time of a charge across a Debye length), the time unit seems absurdly small. Other scalings may provide different perspectives on important phenomena and should be investigated. Similarly, the diffusion length is scaled by the characteristic diffusion coefficient of an ion *inside* the channel. In what follows, we omit the tildes.

The 3D geometry of the ion channel can be well approximated [45, 46] by a reduced 1D problem along the axial direction z, with a cross-sectional area factor The non-dimensional cross-section area $A(z)$ related to the dimensional quantity $A_{dimensional}$ by $A_{dimensional} = \lambda^2 A$. $A(z)$ included as in, e.g. [23, 46, 45, 22, 38, 3, 17, 19, 35, 13, 1], see Figure 2,

$$\frac{1}{A(z)} \frac{d}{dz}(A(z)\phi_z) = z^T c + \rho_0(x), \tag{5}$$

$$\frac{dc_i}{dt} + \frac{1}{A}\frac{d}{dz}J_i = 0, \quad i = 1, \cdots, n, \tag{6}$$

where

$$J_i = -A(z)D_i(z)c_i \frac{d}{dz}\frac{\delta \mathcal{A}}{\delta c_i} = D_i(x)\left(c_{i,z} + c_i\left(z_i\phi_z + \sum_{j=1}^{N} g_{ij}c_{j,z} - \sigma c_{i,zzz}\right)\right). \tag{7}$$

**Table 1** lists the values of the dimensional model parameters considered in this work.

Table 1: Value of dimensional model parameters

| Parameter | Description | Quantity |
|---|---|---|
| $\bar{c}$ | Bulk ionic concentration*a* | 0.1M |
| $\varepsilon$ | Dielectric constant | $78\varepsilon_0 \approx 6.9 \cdot 10^{-10}$ F/m |
| $D_i(0)$ | Diffusion coefficient of species $i$ inside the channel*a* | $0.8 \cdot 10^4$ cm$^2$/s |
| $g_{11}, g_{12}, g_{22}$ | Ion-ion interaction energy parameters | $[14.6, 13.07, 3.64] \cdot 10^{-21}$ J |
| $\sigma$ | Ion-ion high-order interaction energy | $6.08 \cdot 10^{-22}$ J |
| $\lambda$ | Debye length | 0.97nm |
| $\tau$ | Characteristic time | $1.2 \cdot 10^{-18}$s |

*a* assumed equal for all species; Gating is a broad phenomenon, and we do not expect that its study would require a careful choice of parameters. The above parameters were chosen to be within a reasonable magnitude to reflect biological conditions, while remaining in a regime that enables numerical study. In particular, we do not attempt to describe real channels. The non-dimensional problem parameters are taken to be

$$A(z) = 1 + z^2, \quad \rho_{permanent} = c_{permanent}\, e^{-z^4}, \quad D_i(x) = 20(1 - 0.9e^{-z^4}), \quad \varepsilon(z) \equiv 1.$$

Note that from a modeling point of view, the choice of $\rho_{permanent}$, rather than the commonly used piece-wise constant density, is plausibly indifferent. That is to say, it is plausible that the choice of description of the density will have no effect on our results. The numerical problem, however, is easier to solve when $\rho_{permanent}$ is smooth. The boundary conditions are

$$\begin{cases} \phi \to \phi_+, & for \quad z \to \infty, \\ \phi \to 0, & for \quad z \to -\infty, \\ c_i \to \bar{c}_i, & for \quad z \to \pm\infty, \end{cases} \qquad (8)$$

where the electric potential of the left bath is chosen to be the reference potential $\phi = 0$. Since the bath concentrations are equal on both sides, the current through the channel is driven in this simplified model only by applied voltage difference $\phi_+$ between the left and right bathing solutions. The electrochemical potential difference reduces to

$$\mu_i(\infty) - \mu_i(-\infty) = \phi_+ = \left.\frac{\delta \mathcal{A}}{\delta c_i}\right|_{\pm\infty} = J_i \int_{-\infty}^{\infty} \frac{dz}{A(z) D_i(z) c_i(z)}, \qquad (9)$$

where the last equality is obtained by isolating $\frac{\delta \mathcal{A}}{\delta c_i}$ in equation (7).

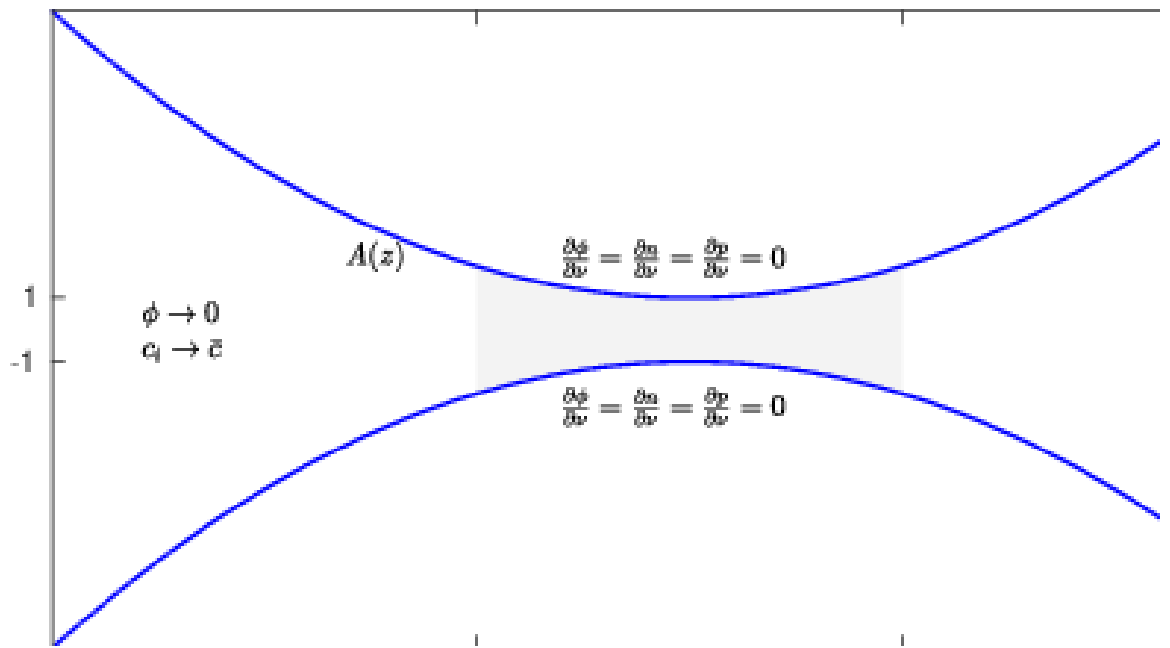

Figure 2: Illustration of model geometry and boundary conditions: z measures distance symmetrically through the channel from the left bath to the right bath. The membrane walls are both flux-free and electrically insulating.

# 3 Steady-state solutions

The steady-state equations satisfy

$$\frac{\delta \mathcal{A}}{\delta c_i} = \lambda_i, \quad i = 1, \cdots, N,$$

where $\lambda_i$ is a Lagrange multiplier associated with charge conservation. The boundary conditions (8) at $z = -\infty$ imply that $\lambda_i = 0$ for $i = 1, \cdots, N$. Furthermore, by (9), the steady-state equations take the form

$$\sigma c_{i,zz} = \log \frac{c_i}{\bar{c}_i} + z_i \phi - \phi_+ \frac{\int_{-\infty}^{z} \frac{ds}{D_i(s) A(s) c_i(s)}}{\int_{-\infty}^{\infty} \frac{ds}{D_i(s) A(s) c_i(s)}} + \sum_{j=1}^{N} g_{ij}(c_j - \bar{c}_j), \quad i = 1, \cdots, N. \tag{10}$$

coupled with Poisson's equation (5).

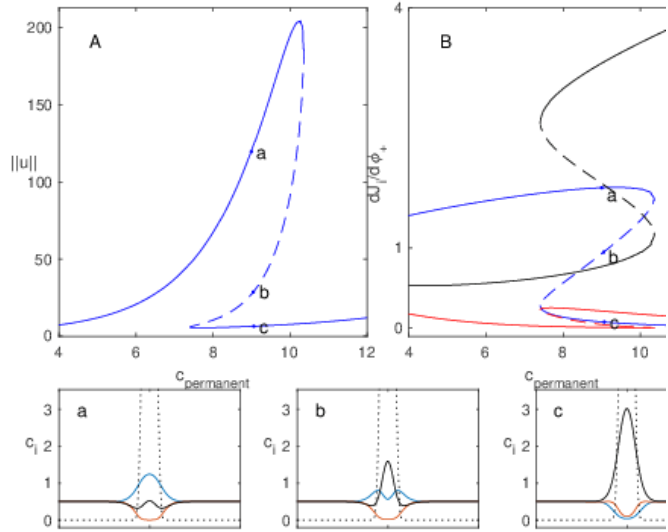

Figure 3: A: The solution norm $\| u \|$, see (11), as a function of $c_{permanent}$. B: $J_1'$ (blue), $J_2'$ (red) and $J_3'$ (black) as a function of $c_{permanent}$. a-c: Solution profiles corresponding to points in A and B. Equation parameters are $\phi_+ = 0$, $z_1 = z_3 = 1$ and $z_2 = -2$, $\sigma = 0.1$, $g_{11} = 2.4$, $g_{12} = 2.15$, $g_{22} = 0.6$, $g_{13} = 2.3$, $g_{23} = 0.2$, $g_{33} = 0.1$. Dotted curve marks the channel region.

To map the steady-state solutions of (2), we solve equations (10) with $\phi_+ = 0$ using continuation with pde2path [50, 6] where the density of permanent charge is the continuation parameter. The computational domain is taken to be $[-L, L]$ where $L = 50$ for which we numerically verify that finite domain effects are negligible (data not shown). The resulting bifurcation diagram of (10) is presented in Figure 3A. Each point on the curve $(c_{permanent}, \| u \|)$ (branch) represents a solution $u$ of (10) with a corresponding parameter $c_{permanent}$, and where the solution norm is defined as

$$\| u \| = \frac{1}{2L} \int_{-L}^{L} w(z)[c_1^2(z) + c_2^{-2}(z) + c_3^2(z) + \phi^2(z)] \, dz, \quad w(z) = sech(5z), \quad (11)$$

in aim to provide a measure that helps differentiate between two steady states on the equation. Solid branches correspond to solutions which are stable with respect to the dynamics of (2), while dashed branches correspond to unstable solutions. Thus, for $c_{permanent} \approx 7.5 - 10$, the system gives rise to multiple steady-states, which lie on two stable branches and one unstable connection branch. The properties of the different branches differ because of the different ionic species that enter the channel in the different cases, Figures 3a-3c. For example, in state 'a', $c_1$ is the dominant species inside the channel, while in state 'c', the other cationic species $c_3$ dominates over $c_1$. The unstable state 'b' reflects a more balanced situation where both cationic species enter the channel.

We had expected that the permanent charge would be roughly balanced by mobile charge inside the channel domain. We often find, however that the permanent charge is also balanced by lobes of mobile charges at the channel ends.

It is the competition between two cationic species in the channel region that gives rise to multiple steady-states. In the case of a single cation species (counter balanced by an anion), there is no competition, and accordingly we do not observe multiple steady-states in the system of study (data not shown).

To identify the opened and closed states, it is instructive to monitor the ionic flux of the solutions along the branches. The flux of ionic species $c_i$ is given by, see (9),

$$J_i = \frac{\phi_+}{\int_{-\infty}^{\infty} \frac{ds}{D_i(s)A(s)c_i(s)}}.$$

Here, however, we consider the case of zero applied voltage $\phi_+ = 0$ where all solutions correspond to zero ionic flux $J_i = 0$. Accordingly, in Figure 3B, we monitor the flux derivatives (slope 'conductance') at $\phi_+ = 0$

$$\left.\frac{dJ_i}{d\phi_+}\right|_{\phi_+=0} = \frac{1}{\int_{-\infty}^{\infty} \frac{ds}{D_i(s)A(s)c_i(s)}},$$

which is proportional to the ionic fluxes for small enough applied voltage, $|\phi_+| \ll 1$. Particularly, we observe that for $|\phi_+| \ll 1$, state 'a' is a conductive state for species $c_1$, but sub-conductive for species $c_3$, while state 'c' is a sub-conductive or closed state for species $c_1$ but conductive for species $c_3$. The negative correlation between the conductance of the cationic species is an implication of the competition between them.

## 4 Bi-stability and gating via a hysteresis loop

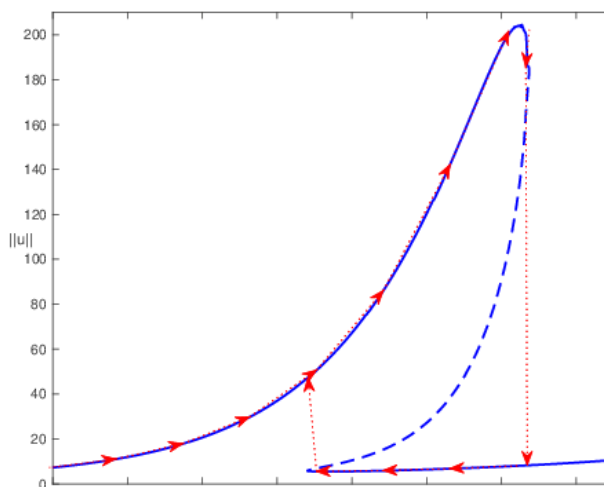

Figure 4: Hysteresis loop in system of Figure 3.

The sub-critical bifurcation in Figure 3A suggests that the system undergoes hysteresis as the permanent charge varies, see illustration in Figure 4. Hysteresis suggests a mechanism of single channel gating. For example, if the system is at point 'a' and driven toward higher permanent charge, the solution will initially follow its branch with $J_1'(\phi_+) \approx 1.75$, see Figure 3B. If permanent charge density, however, exceeds the bifurcation point, the solution will switch to the lower branch with $J_1'(\phi_+ = 0) \approx 0.033$. Significantly, however, the bifurcation diagram and the hysteresis loop, see Figures 3A and 4, describe only steady-state solutions of (2) rather than transients. Therefore, gating achieved by tracing the hysteresis loop as in Figure 4 requires varying the permanent charge coefficient $c_{permanent}$ much more slowly than the relaxation time of (2) so that the system remains close to equilibrium. *Thus, while hysteresis implies a mechanism of gating under slow changes in the system parameters, it is not clear whether gating would occur for faster changes, e.g., in noisy environments*.

To study current dynamics under fast changes, we first conduct a simulation of the steric PNP-CH (2) The computational domain is taken to be [-50 50], which is verified to be large enough so that finite domain effects are neglibile. in which the permanent charge is decreased and increased for relatively short times during the simulation, see Figure 5B. As expected, as the permanent charge density is pushed beyond the bifurcation points for a sufficient period of time, the solution switches to a different branch, leading to a new current level, see Figure 5A.

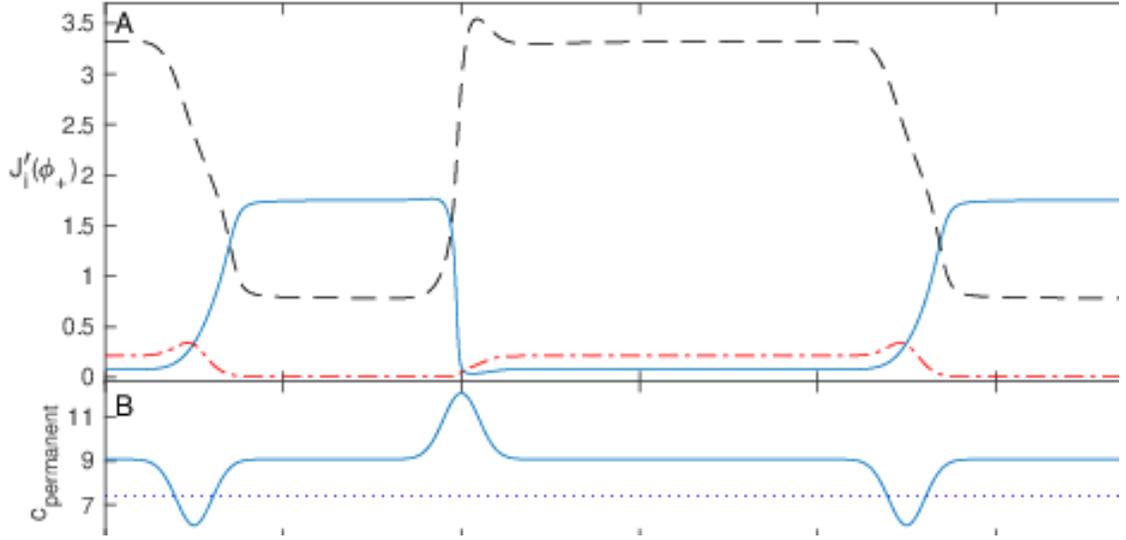

Figure 5: A: Slope 'conductance' $J'_i(\phi_+ = 0)$ as a function of time for $i = 1$ (solid), $i = 2$ (dash-dotted) and $i = 3$ (dashed). B: The permanent charge density coefficient $c_{permanent}$ as a function of time. The bifurcation point $c_{permanent}^{bif} \approx 7.4$ is marked by a dotted line. Equation parameters are as in Figure 3.

The resulting current vs time graph resembles (but see subsequent section for a detailed comparison) the experimental measurements of gating, see Figure 1. The distinct current levels observed correspond to different equilibrium solutions. Accordingly, the switching time between the different current level correspond to the relaxation time to the new equilibrium state. We observe that the switching time is slower than the characteristic time scale of the system, and depends on the direction of the transition. For example, the switching from current level $J'_3 \approx 0.78$ to $J'_3 \approx 3.32$ (see dashed curve in Figure 5A) occurs in roughly 20 units of time, i.e., 20 times slower than the characteristic time of ionic diffusion in the channel. In general, *to allow a system to relax to an equilibrium state, the noise in the system must be sufficiently small over the full relaxation time*. Therefore, it is unlikely that single channel gating would be observed in the presence of noise, in the the system we are studying.

To demonstrate this, we solve the steric PNP-CH (2) where Gaussian noiseGenerated by Comsol's random function with normal distribution. is introduced to the permanent charge density coefficient, see Figure 6B. The permanent charge density coefficient $c_{permanent}$ frequently crosses the bifurcation point (marked by a dotted line in Figure 6B), but remains below the bifurcation point for brief periods of time which are typically much shorter than the relaxation time of the system. Therefore, as expected, no switching behavior is observed for the full simulation time, see Figure 6A.

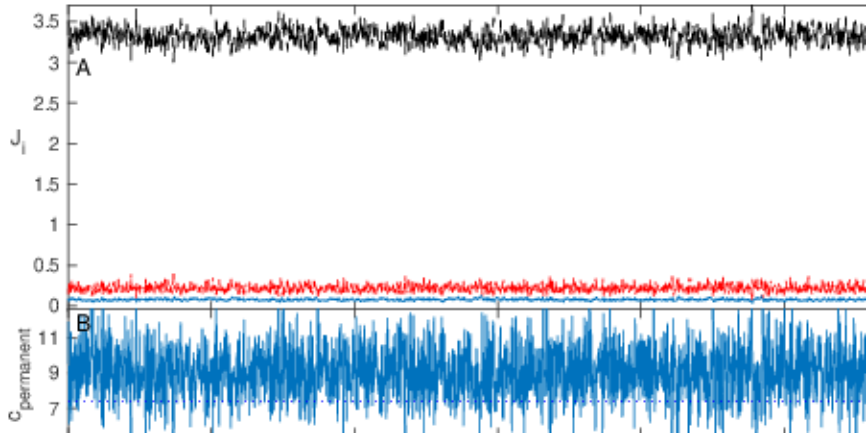

Figure 6: A: Flux derivative $J'_i(\phi_+ = 0)$ as a function of time for $i = 1$ (solid), $i = 2$ (dash-dotted) and $i = 3$ (dashed). B: The permanent charge density coefficient $c_{permanent}$ as a function of time. The bifurcation point $c_{permanent}^{bif} \approx 7.4$ is marked by a dotted line. Equation parameters are as in Figure 3.

The numerical study presented in Figures 5 and 6 was confined to the case of zero applied voltage $\phi_+$, which is easier to compute and analyze. To show that our results are applicable to more general condition, we also present a simulation with $\phi_+ = 1$, see Figure 7.

As expected, the applied voltage does not change the qualitative behavior of the system.

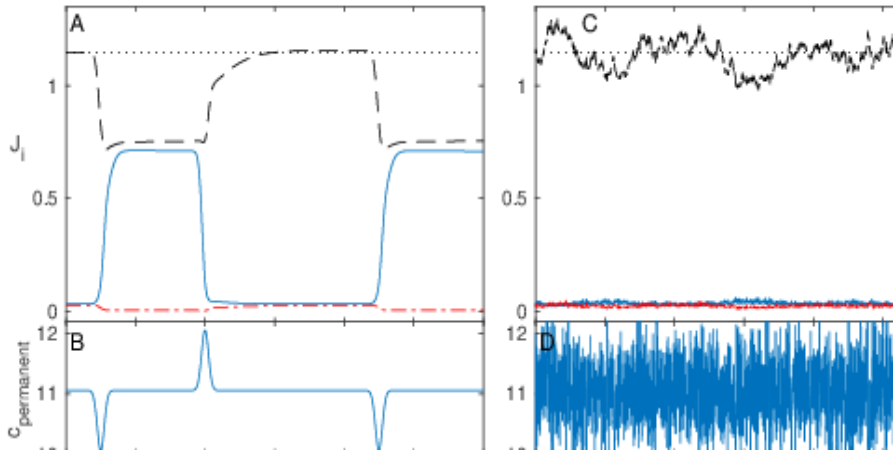

Figure 7: Simulation of the steric PNP-CH (2) with $\phi_+ = 1$ and $\sigma = 5 \cdot 10^{-3}$. Rest of the equation parameters are as in Figure 3. A, C: Fluxes $J_i$ as a function of time for $i = 1$ (solid), $i = 2$ (dash-dotted) and $i = 3$ (dashed) where the corresponding permanent charge density

coefficient $c_{permanent}$ as a function of time is presented B and D, respectively. The bifurcation point $c_{permanent}^{bif} \approx 7.4$ is marked by a dotted line.

All simulations of the steric PNP-CH (2) are conducted using Comsol 5.3.

## 5   Relevance to gating phenomena in biological channels

We have shown that the steric PNP-CH model (2) gives rise to switching behavior, albeit under customized conditions, and in the absence of noise. These conditions are so different from those in biological channels that we conclude our model does not describe the spontaneous gating of single channels observed experimentally. Indeed, ionic channels are exposed to thermal fluctuations which can be very significant in such a nano-scaled system as is obvious in any simulation of molecular dynamics, particularly if one remembers the strength of Coulomb's law. Therefore, ionic channels operate in a fluctuating noisy environment of large magnitude, where noise is manifested in various ways including the distribution of fixed charge, domain of the ion channel, location of the free ions themselves, etc.

One of the goals of this paper is to define the problem of interest, and set up criteria that will allow future work to determine whether an observed switching behavior begins to describe the spontaneous gating of single channels. Accordingly, we now ask whether the switching behavior of the steric PNP-CH model, observed, e.g., in Figure 5, captures the essential features of gating in biological ion channels. The steric PNP-CH switching mechanism has the following characteristics:

1. Switching occurs due to transitions between multiple steady states. The characteristic duration of switching is determined by the relaxation time of the system, which, as expected, is observed to be much longer than the characteristic ionic transport time in the system.

2. Switching is induced when an effective parameter crosses a critical value (bifurcation point) for a sufficient duration of time. This duration of time is tightly related to the relaxation time of the system.

3. Multiplicity of steady states is observed in narrow parameter regimes.

4. Opened channel and zero (sub-conductance) currents are not sharply determined by the system. We observed, for example, that noise in $c_{permanent}$ leads to comparable noise in the resulting currents through the channel.

5. Multiplicity of steady states stem from the competition between multiple cationic species in the channel region. Consequently, a system with a single cationic species will not give rise to multiple solutions, and will not be able to describe gating via the mechanism of study.

6. Since each steady-state represents a different balance between cationic species in the channel region, correlation between the fluxes of different cationic species are observed. For example, upon switching, the channel may 'close' for Sodium ions but at the same time would 'open' for Potassium ions. While this kind of behavior is not what we seek here, it might be of great importance in the sequential changes in selectivity that define transporters vs. ion channels.   Reference [31] gives a taste of this complexity in a modern physical context.

The above features are at odds with the characteristics of gating in biological ion channels. Indeed, gating in channel is a generic and wide-spread phenomena. It can occur in a channel dominated by single cationic species [21, 53, 36], and therefore it is not a consequence of competition between multiple cationic species. Currents are either (nearly) zero or at a definite level, independent of time and surprisingly insensitive to the vast thermal fluctuations that change the location of charges in the channel protein and its pore by substantial amounts, once the channel is open.

Future work addressing this problem must take into account the defining properties of gating in single channels:

1. Gating is universal, observed in many types of channels with very different structures.

2. The sudden (~1 microsecond) switching from nearly zero to a definite value, often between 10 and 1000 pa.    Temporal resolution of the switching event at a 100 nano-second time scale reveals stochastic variations in pore current of great diversity, including graded, stepwise and oscillatory variations [37, 48].

3. The rectangular nature of the wave form. The amplitude of the single channel current is independent of time, once the channel is open and independent of the duration of the opening from some 50 microseconds to even tens of seconds in favorable experimental situations.

4. Closed and open duration histograms often can be fit very well (but not perfectly) with single exponential functions.

## 6  Discussion

In this study, we have addressed the question whether gating occurs due to noise-induced fast transitions between multiple steady (equilibrium) states on an ion channel. We have considered this question by studying the switching behavior in PNP equations with high-order steric effects. We have shown that for two cationic species, the equation does produce switching between multiple solutions for rather narrow ranges of ion-ion interactions parameters and permanent charge densities in the channel region, and in the absence of noise. The observed switching behavior, however, does not have the essential defining properties of gating in biological channels.

The noise we consider is a rapid fluctuation, independent of its history (similar to white noise), in the effective permanent charge density (residue) concentration. Importantly, rapid means much faster than the relaxation time of the model. A key observation of this study is that noise does not induce switching,   at least of the type studied here. Indeed, to allow a system to relax to a new equilibrium state, the noise needs to exceed a certain threshold and remain in some range above this threshold over the full relaxation time of the model. This, however, is an unlikely eventTo say the least. Consider, for example, a normally distributed noise $\xi \sim N(0,\sigma)$ that fluctuates at an atomic time scale of $10^{-15}$ seconds. The probability that this signal will exceed $\sigma/5$ for a period of $10^{-10}$ seconds is less than $10^{-100}$.. This observation is of general nature. Indeed, in over-damped models (gradient flow), the relaxation time is nearly always longer than the characteristic time of change in the model. Therefore, it is unlikely that noise will induce gating in such models. Of course, we can only discuss noise of the type and structures we have computed. Channels have larger structures, both long and short lasting, that might move cooperatively and create single channel gating. Indeed, the conformation changes of classical biophysics are of this type, and deserve to be studied in physical representations that are more realistic than traditional chemical kinetics.

The observed switching behavior reflects a competition between multiple cationic species, such that each steady-state represents a different balance between cationic species in the channel regions. This kind of behavior is not characteristic of gating. It has, however, an important characteristic found in transporters, that have structure similar to channels or branched channels. Transporters allow different ions to flow in their different states and in that sense have state coupled selectivity. The switching behavior observed here also has different selectivity in different states because each state has a different balance between cationic species. We will present a study of transporters this in further publications.

Gating is a universal phenomena, observed in many types of channels with such different structures, and therefore should be described in rather generic models of ion channels, we suppose. In accordance, we focus in this work on electrolyte structure and dynamics and do not make a careful choice of parameters, take into account protein specific details, nor protein response (polarization). Specifically, our study focuses on the 1D-reduced version of the steric PNP-CH model since it has been successful in predicting permeability and selectivity of ionic channels in their open state, and since it gives rise to multiple steady states. Within the steric PNP-CH model, we further assume a simplified geometry, uniform fixed charge distribution within the channel, and consider effective ion-ion interaction parameters that give rise to multiple steady statesWhile the theory relating parameters $g_{ij}$ to ionic radii is available [23], to the best of our knowledge, no such theory had been developed for the higher order parameter $\sigma$.. Thus, our model is a phenomenological representation of an ion channel. It is important to realize that all models and simulations of ion channels are phenomenological. For example, many of the hundreds of parameters used in simulations of molecular dynamics are determined from macroscopic measurements in spatially homogeneous systems (as they should be in our opinion, we hasten to add). Simulations will thus be phenomenological until scientists learn how to compute the parameters of molecular dynamics a priori from quantum mechanical analysis of electrolyte solutions. Even then the question will remain whether the parameters are appropriate for the special conditions within an ionic channel, with its $>10M$ ionic strength. These issues become particularly vivid when we realize the sensitivity of biological results (of selectivity, for example) to the choice of combining rules for atoms of unequal charge or properties.

In this study, we have considered Gaussian noise. Given the definite picture arising from this study, we do not believe that other types of noise will induce gating or change the qualitative picture, and accordingly we have not extended our study to consider noise with different characteristics. We do not believe, on the other side, that Gaussian noise describes accurately the noise environment in a channel. Rather, it is plausible that various charged and non-charged components of the channel would fluctuate in a cooperative, coordinated, motion giving rise to unique noise characteristics.

The open question of gating is interlinked with an open question regarding the robustness of ion channels: A biological ion channel is subject to enormous noise, and thus it seems more than implausible that its outputs are well-defined currents which are independent of time, as pointed out long ago [9]. Yet, this is the experimental picture - the observed currents of biological ion channels reflect only a tiny part of the noise in the system (while they are opened or closed). Thus, it seems necessary to imagine a sort of 'eigenstate' in which current can flow only when the channel has particular spatial distributions of mobile and permanent charge, i.e., when the channel has particular conformations of charge (not mass), and these produce potential profiles that allow current flow. The existence of Coulomb blockade in simulations of calcium and sodium in channels [26, 27] supports this view. We will address these issues in further publications.

## Acknowledgments

This research was supported by grant #2016106 from the United States – Israel Binational Science Foundation (BSF). The author thanks Hannes Uecker for valuable advice in the application of pde2path.